\documentclass[aps, prd, twocolumn, superscriptaddress,nofootinbib]{revtex4}
\usepackage{graphicx}
\usepackage{dcolumn}
\usepackage{amssymb,amsmath,amsthm,mathrsfs}

\begin{document}

\newcommand{\Eq}[1]{Eq. \ref{eqn:#1}}
\newcommand{\Fig}[1]{Fig. \ref{fig:#1}}
\newcommand{\Sec}[1]{Sec. \ref{sec:#1}}

\newcommand{\PHI}{\phi}
\newcommand{\vect}[1]{\mathbf{#1}}
\newcommand{\Del}{\nabla}
\newcommand{\unit}[1]{\mathrm{#1}}
\newcommand{\x}{\vect{x}}
\newcommand{\ScS}{\scriptstyle}
\newcommand{\ScScS}{\scriptscriptstyle}
\newcommand{\xplus}[1]{\vect{x}\!\ScScS{+}\!\ScS\vect{#1}}
\newcommand{\xminus}[1]{\vect{x}\!\ScScS{-}\!\ScS\vect{#1}}
\newcommand{\diff}{\mathrm{d}}

\newcommand{\be}{\begin{equation}}
\newcommand{\ee}{\end{equation}}
\newcommand{\bea}{\begin{eqnarray}}
\newcommand{\eea}{\end{eqnarray}}
\newcommand{\vu}{{\mathbf u}}
\newcommand{\ve}{{\mathbf e}}
\newcommand{\vk}{{\mathbf k}}
\newcommand{\vx}{{\mathbf x}}
\newcommand{\vy}{{\mathbf y}}

\newcommand{\uden}{\underset{\widetilde{}}}
\newcommand{\den}{\overset{\widetilde{}}}
\newcommand{\denep}{\underset{\widetilde{}}{\epsilon}}

\newcommand{\nn}{\nonumber \\}
\newcommand{\dd}{\diff}
\newcommand{\fr}{\frac}
\newcommand{\del}{\partial}
\newcommand{\eps}{\epsilon}
\newcommand\CS{\mathcal{C}}

\def\be{\begin{equation}}
\def\ee{\end{equation}}
\def\ben{\begin{equation*}}
\def\een{\end{equation*}}
\def\bea{\begin{eqnarray}}
\def\eea{\end{eqnarray}}
\def\bal{\begin{align}}
\def\eal{\end{align}}


\title{Inflationary tensor fluctuations, as viewed by Ashtekar variables and their
imaginary friends}

\newcommand{\addressImperial}{Theoretical Physics, Blackett Laboratory, Imperial College, London, SW7 2BZ, United Kingdom}

\author{Laura Bethke}

\author{Jo\~{a}o Magueijo}
\affiliation{\addressImperial}


\date{}

\begin{abstract}
We investigate tensor modes in inflationary scenarios from the
point of view of Ashtekar variables and their generalizations
labelled by Immirzi parameter $\gamma$, which we'll assume
imaginary.
Being careful to properly define the classical perturbed Hamiltonian by
taking several subtleties into account, we reproduce, on-shell,
the usual expression found in cosmological perturbation theory.
However the quantum Hamiltonian displays significant
differences, namely in the vacuum energy and fluctuations of the
various modes. We can identify combinations
of metric and connection variables representing graviton states, noting that
before reality conditions are imposed there are gravitons and anti-gravitons. It turns out
that half of these modes have negative energy but after defining the inner product we conclude
that they are non-physical and should be selected out. We are left with the
usual graviton modes but with a chiral asymmetry in the the vacuum
energy and fluctuations. The latter depends on $\gamma$ and on the
ordering prescription (namely in the Hamiltonian constraint). Such
an effect would leave a distinctive imprint in the polarization of
the cosmic microwave background, thus finally engaging quantum
gravity in meaningful experimental test.
\end{abstract}



\maketitle


\section{Introduction}
Inflation~\cite{inflation} is an impressive sociological
achievement. Yet, purely on scientific merits, its predictive
value has been questioned, with the epithet ``theory of anything''
being sometimes applied to its vast array of models. This is not
entirely fair, considering that, at least within the most
``reasonable'' models, a number of consistency conditions must be
satisfied, notably those relating the spectral index of scalar
fluctuations and the amplitude of tensor modes. Should
gravitational waves be detected, say via the polarization of the
CMB, and should they satisfy these consistency
conditions, it would be hard to deny the predictive power of
inflation.

And yet, from the point of view of theories of quantum gravity,
the inflationary {\it quantum} mechanism for generating tensor
fluctuations inevitably raises concerns. Strictly speaking, we
should be in possession of the full theory of quantum gravity to
understand graviton modes even at the perturbative level. One
might argue that inflationary fluctuations are protected from the
detailed workings of quantum gravity. But delving deeper into the
issue reveals that details do matter. A re-examination of
inflationary tensor modes from the standpoint of Ashtekar's
formalism leads to the prediction of a chiral gravitational wave
background~\cite{prl}, unlike in the standard inflationary
calculation, based on the second order formalism. This may be a
curse or a blessing. From the point of view of quantum gravity it
certainly is cause for celebration: it might open up the field to
observation, a basic requirement for any truly scientific theory.
It is remarkable that it has occasionally been claimed that the
problem of quantum gravity has been solved, when no experimental
predictions have been made, let alone verified.

In this paper we provide the details behind our recent {\it
Letter}~\cite{prl}, where the chirality of the perturbative vacuum
of quantum gravity using Ashtekar variables was derived. The Ashtekar formalism constitutes
a promising scheme for quantizing the gravitational
field~\cite{ashbook,pulbook,rovbook,thbook}. At its core lies the
idea that the connection (or its holonomies), rather than the
metric, should be the central gravitational variable driving
quantization: tools employed to quantize non-perturbative gauge
theories, such as the Wilson loop, can then be used leading to
great progress. Furthermore the formalism relies on the
application of a canonical transformation (dependent on Immirzi
parameter $\gamma$) upon the Palatini-Kibble
spin-connection~\cite{thbook}. For $\gamma=\pm i$ the connection
becomes self-dual (SD) or anti-self-dual (ASD), leading to a
number of nice mathematical properties and simplifications.
The parameter $\gamma$, however, is usually left undefined. It
drops out of the classical field equations but is expected to
leave a quantum imprint, similar to the theta parameter in QCD.

Specifically, we showed in~\cite{prl} that the quantum vacuum
energy and fluctuations predicted by the theory display a
chirality dependent on $\gamma$. For simplicity, we
laid out the argument for the extreme cases $\gamma=\pm i$ and
then just presented the answer for a general $\gamma$. Here we
provide the detailed calculation. The strategy followed in our
work is simple: we never stray far from standard cosmological
perturbation theory~\cite{muk,lid}. Well established {\it
classical} results in Cosmology must obviously have exactly
equivalent descriptions in Ashtekar's formalism; if they don't
something has gone awry. Furthermore, the Ashtekar quantization
procedure should be mapped, in some approximation, onto the usual
inflationary calculation of tensor vacuum quantum fluctuations. If
experimentally meaningful differences arise, one should cherish
them, but also understand their origin. Thus, in this paper we
shall present two types of results: classical and quantum. Whereas
classically we will just rediscover well-known results within a
different formalism, the quantum mechanics will contain true
novelties.

There is a popular belief that cosmological perturbation theory in
Ashtekar's formalism ``has all been done before'' or ``is just an
exercise for the student''. Nothing could be more far-removed from
reality. The exercise is highly nontrivial, particularly the
Hamiltonian solution. Even within the second order
formalism~\cite{deser}, a Hamiltonian recast of cosmological
perturbation theory is far from straightforward (although it has
indeed been done before~\cite{langlois}). As for the first order
formalism, apart from partial results obtained in the context of
loop quantum cosmology~\cite{lqc} (with a different slant, and
therefore bypassing important issues), pitifully little has been
done.

In Sections~\ref{classical},~\ref{realitytorsion} and~\ref{secham} we
will try to patch this ungainly hole. In Section~\ref{classical}
we solve the Einstein-Cartan equations to find the classical
solution for tensor perturbations in a de Sitter Universe in terms
of Ashtekar variables. Illuminating insights on duality and
helicity are obtained, clearing up some misconceptions found in the literature.
Also, the issue of the reality conditions will be clarified (Section~\ref{realitytorsion}).
We then proceed (Section~\ref{secham})  to
rediscover our results within the Hamiltonian formalism, essential
for canonical quantization. As we will see, one does {\it not}
recover the results of standard cosmological perturbation theory,
unless a number of subtleties are taken into account.

The second part of this paper is concerned with the quantization
of this system. In Section~\ref{coms} we set up the quantization
procedure by obtaining the commutation relations in terms of a
mode expansion. Graviton states can then be identified by
examining the structure of the Hamiltonian (Section~\ref{quantum
ham}). In general we find twice as many modes as expected. However
the reality conditions impose a specific form for the inner
product, which renders half of these modes unphysical. These are
also the modes which display negative energy and which do not
exist classically. If we restrict ourselves to physical modes we
therefore recover the standard gravitons; however the vacuum
fluctuations and energy turn out to be chiral, establishing the
main result in this paper. In Section~\ref{vacfluc} we explain how
this chirality may differ when vacuum fluctuations and vacuum
energy are contrasted, and how the result depends on the
ordering prescription. Finally, in a concluding Section we
summarize our results and outline plans for future work.

Throughout this paper we shall use units for which $\hbar=c=1$ and
we parameterize the strength of gravity with $l_P^2=8\pi G$. We'll
be concerned with the real world, so the metric will invariably be
Lorentzian; its signature is taken to be $-+++$.

\section{Classical solution: helicity vs duality}
\label{classical} We will first map well-known results pertaining
cosmological tensor perturbations into the Ashtekar formalism. A
prominent issue is that of relating helicity states (right and
left handed) and duality states (self-dual, SD, and
anti-self-dual, ASD). Even though the issue was cleared long
ago~\cite{ash0}, the myth has persisted that these two types of
states align: the right handed graviton is SD and the left handed
one ASD. Instead, it should be obvious that this cannot be true:
the two types of states can never align because helicity
states are real whereas duality states must be complex for a
Lorentzian space. Reality conditions therefore relate SD and ASD
states; but they can never impose a constraint upon helicity
states.

A close analysis reveals that a proper understanding of the
relation between these two types of states can only be reached by
including positive and negative frequencies into expansions. This
simple technical point was made in~\cite{ash0} but missed
in much of the subsequent literature
(e.g.~\cite{gravitons,leelaur}). It is the cause of much confusion
as well as a few paradoxes. For example, the negative energy modes
found by Witten~\cite{wittenym}, associated with perturbations of the
Kodama state~\cite{kodama}, can never be helicity modes in a Lorentzian
space, as explained later in this paper.

\subsection{Conventions and the background solution}
Let the unperturbed Universe be de Sitter space-time foliated
using a flat slicing. Using comoving spatial coordinates and
conformal time the background metric is therefore: \be
ds^2=a^2(-d\eta^2 +dx^2+dy^2+dz^2) \ee with \be a=-\frac{1}{H\eta}
\ee where  $H^2=\Lambda/3$ and $\eta<0$. The tetrad basis is
$e^I_\mu=a\delta^I_\mu$ and the non-zero connection forms are
$\Gamma^i_{\;0}=He^i$, where $i=1,2,3$ are 3D Lie algebra indices
(as usual~\cite{rovbook,thbook}, we use $I$ and $\mu$ for 4D group
and space-time indices, respectively). From these we can find the
densitized inverse triad, defined as \be E^a_i=\det\left(e^j_b\right)e^a_i \ee and the self-dual connection.

We shall use the following conventions\footnote{We
follow~\cite{thbook}, which is just about the only reference where
the author deigned to check the consistency of some of the conventions.}.
We dualize with a Levi-Civita symbol derived from
$\epsilon_{0123}=1$ (as opposed to $\epsilon^{0123}=1$) and define
the SD connection so that $\star A^i=+i A^i$. Furthemore we map
anti-symmetric rank-2 spatial indices into vectors via
 $\Gamma^{ij}=\epsilon^{ikj}\Gamma^k$ (see~\cite{thbook}, pp.127),
or equivalently \be\label{gammaconv}
\Gamma^i=-\frac{1}{2}\epsilon^{ijk}\Gamma^{jk}\; . \ee (This
convention renders the spatial Cartan equation as $De^i=de^i
+\epsilon^{ijk}\Gamma^j\wedge e^k$). Then the SD connection can be
defined by: \be\label{sddef} A^i=\Gamma^i+\gamma \Gamma^{0i} \ee
with $\gamma=i$; indeed it is straightforward to check that with
our conventions $\star A^i=+iA^i$. The ASD connection ($\star
A^i=-i A^i$) follows from $\gamma=-i$, the Immirzi connection from
a general complex $\gamma$, and the Barbero connection from a real
$\gamma$.

Therefore, the background (unperturbed) space has densitized inverse triad and
self-dual connection given by: \bea
E^a_i&=&a^2\delta^a_i\label{zeroth1}\\
A^i&=&-\frac{1}{2}\epsilon ^{ijk}\Gamma^{jk} +i\Gamma^{0i} =iH
e^i\label{zeroth} \eea where $a=1,2,3$ are spatial indices for the
base manifold. It can be easily shown that these satisfy: \be
B^a_i+H^2 E^a_i=0\;  \ee where $B^a_i$ is the $SU(2)$ magnetic
field of $A^a_i$. For a more general $\gamma$ we have \be
A^i=\gamma H e^i\ee leading to \be B^a_i- \gamma^2 H^2 E^a_i=0\;
.\ee

\subsection{Perturbation variables}
We want to study tensor perturbations around this background
solution. It is standard~\cite{muk,MTW} to write them as:
\be\label{pertmet} ds^2=a^2[-d\eta^2
+(\delta_{ab}+h_{ab})dx^adx^b] \ee where $h_{ab}$ is a symmetric, TT
(transverse and traceless) Cartesian tensor. The associated triad
is therefore perturbed as: \be e^i_a
=a{\left(\delta^i_a+\frac{1}{2}h^i_a\right)} \ee so that for the
inverse triad and densitized inverse triad we have: \bea
e^a_i&=&\frac{1}{a}{\left(\delta^a_i-\frac{1}{2}h^a_i\right)}\\
E^a_i&=&a^2{\left(\delta^a_i -\frac{1}{2}h^a_i\right)}\; , \eea
where we have raised and lowered indices in $h_{ab}$ with a
Kroenecker-$\delta$ mixing $a$ and $i$ types of indices (i.e.
space-time and algebra indices). More generally, to leading order
we may confuse the $a$ and $i$ indices in all perturbation
variables, lowering and raising them likewise. Throughout this
paper we'll adopt the following convention: we define $\delta
e^i_a$ via the triad \be e^i_a=a\delta^i_a+\delta e^i_a\; .\ee We
then raise and lower indices in all tensors with the
Kronecker-$\delta$, possibly mixing group and spatial indices.
This simplifies the notation and is unambiguous if it's understood
that $\delta e$ is originally the perturbation in the triad. With
these conventions we therefore have:  \bea
e^a_i&=&\frac{1}{a}\delta^a_i -\frac{1}{a^2}\delta e^a_i\\
E^a_i&=&a^2\delta^a_i - a\delta e^a_i \label{pertE}\; .\eea We
could have adopted any other convention but it
turns out that in our case $\delta e_{ij}$ is
proportional to the ``$v$'' variable beloved by
cosmologists~\cite{muk,lid} (see also
Appendix I). For the connection we write: \be\label{pertA}
A^i_a=\gamma Ha \delta^i_a + \frac{a^i_a}{a} \ \ee

As in the usual cosmological treatment we now subject the
perturbations to Fourier and polarization expansions; however the
Ashtekar formalism presents us with some subtleties. The main issues are:
\begin{itemize}
\item If reality conditions are yet to be enforced there must be
graviton and anti-graviton modes, so it's essential not to forget
the negative frequencies in all expansions, and ensure that they
are initially independent of the positive frequencies.

\item For a clearer physical picture, it is convenient to use the
quantum field theory convention stipulating that for free modes
the spatial vector $\vk$ points in the direction of propagation
{\it for both positive and negative frequencies}. This is a simple
point, but spurious couplings between $\vk$ and $-\vk$ modes
otherwise come about, e.g. reality conditions constrain gravitons
moving in opposite directions~\cite{gravitons,leelaur}, which is
physically nonsensical.

\item If the above is employed, the physical Hamiltonian also
should not contain couplings between $\vk$ and $-\vk$ modes
inside the horizon. The presence of such couplings in
the formalism~\cite{gravitons,leelaur} merely reflects not having
properly identified the direction of propagation (and thus the
polarization). As the modes leave the horizon, couplings between
$\vk$ and $-\vk$ may appear, and represent the production of
particle pairs by the gravitational field (where the particles in each
pair move in opposite directions)~\cite{grishchuk}.
\end{itemize}
Bearing this in mind we thus write: \bea \delta e_{ij}&=&\int
\frac{d^3 k}{(2\pi)^{\frac{3}{2}}} \sum_{r}
\epsilon^r_{ij}({\mathbf k}) {\tilde\Psi}_e(\vk,\eta)e_{r+}(\vk)
\nonumber\\
&&
+\epsilon^{r\star}_{ij}({\mathbf k}) {\tilde\Psi}_e^\star (\vk,\eta)
e^{\dagger}_{r-}(\vk)\nonumber\\
a_{ij}&=& \int \frac{d^3 k}{(2\pi)^{\frac{3}{2}}} \sum_{r}
\epsilon^r_{ij}({\mathbf k}) {\tilde\Psi}_{a}^{r+}(\vk,\eta)a_{r+}(\vk)
\nonumber\\
&& +\epsilon^{r\star}_{ij}({\mathbf k}) {\tilde\Psi}_{a}^{r-
\star} (\vk,\eta)a^{\dagger}_{r-}(\vk) \label{fourrier}\eea where,
in contrast with previous literature
(e.g.~\cite{gravitons,leelaur}), $e_{rp}$ and $a_{rp}$  have two
indices: $r=\pm 1$ for right and left helicities, and $p$ for
graviton ($p=1 $) and anti-graviton ($p=-1$) modes. In a frame
with direction $i=1$ aligned with $\vk$ the polarization tensors
are~\cite{MTW}:
\begin{eqnarray}
\epsilon^{(r)}_{ij}&=&\frac{1}{\sqrt2} \left(\begin{array}{ccc}
0&0&0\\
0&1&\pm i\\
0&\pm i&-1\end{array}\right)\; .
\end{eqnarray}
The base functions have form \be{\tilde
\Psi}(\vk,\eta)=\Psi(k,\eta) e^{i\vk\cdot \vx} \ee and we impose
boundary conditions \be\label{ketabig}\Psi(k,\eta)\sim{e^{-i k\eta}}\ee when
$|k\eta|\gg 1$ for both $+\vk$ and $-\vk$ directions~\footnote{We
stress that everywhere in this paper $k=|\vk|>0$.}. Only then does
$\vk$ point in the direction of propagation, as required. This
convention has the essential advantage of identifying the proper
physical polarization: until we know in which sense the mode is
moving we cannot assign to it a physical polarization. The
functions $\Psi_e$ and $\Psi_a$ can in principle be anything
if we allow the amplitudes $e_{rp}$ and $a_{rp}$ to have the necessary time
dependence. However, we may choose $\Psi$ functions
so that they carry the full time
dependence and the $e_{rp}$ and $a_{rp}$ are constant.
Hamilton's equations then merely confirm the latter,  but ${\tilde\Psi}_a^{rp}$ should have
both $r$ and $p$ dependence. In these expansions we have already
selected the physical degrees of freedom, i.e. the Gauss and
diffeomorphism constraints have been implemented to linear order.
Whether or not this is good enough will be commented upon later.

\subsection{Reading off the classical solution}
In order to canonically quantize the theory we need its
Hamiltonian formulation. We'll do this in detail in
Section~\ref{secham}, but stress that we can read off the answer
from cosmological perturbation theory (see Appendix I and
also~\cite{muk,lid}). Indeed the solution presented in Appendix I
is equivalent to solving the problem in the second order
formalism. Plugging it into the expressions for the Ashtekar
connection and imposing a torsion-free condition for relating
metric and connection is equivalent to solving the Lagrange
equations for the Holst action \be \label{holst} S=-\frac{1}{2l_P^2} \int \Sigma^{IJ}\wedge
\left(F_{IJ}+\frac{1}{\gamma}{}^\ast F_{IJ}\right) \, .\ee
Bearing this in mind, we can conclude that functions $\Psi_e$ must
satisfy the same equation as the variable ``$v$'' used by
cosmologists. We have that $\delta e_{ij}=a h_{ij}/2$, so $v\sim
\delta e_{ij}$. Therefore, in a de Sitter background:
\be\label{eqnpsie}
\Psi_e''+{\left(k^2-\frac{2}{\eta^2}\right)}\Psi_e=0 ,\ee
 where $'$ denotes differentiation with respect to conformal time. This has
solution: \be\label{psie} \Psi_e=\frac{e^{-ik \eta}}{2 \sqrt  { k}
} {\left( 1-\frac{i}{k\eta} \right)}\; ,\ee where the
normalization ensures that the amplitudes $e_{rp}$  become
annihilation operators upon quantization. In addition, connection
and metric are related by the torsion free condition:
\be\label{cartan} T^I=d e^I + \Gamma^I_J\wedge e^J=0\; . \ee This
is solved by \bea
\delta\Gamma^0_{\; i}&=& \frac {1}{a}{\delta e'_{ij}} \, dx^j \\
\delta \Gamma_{ki}&=&-\frac{2}{a}\partial_{[k}\delta e_{i ] j}\,
dx^j\; . \eea With the conventions given above the second of these
equations implies: \be \label{deltagamma}\delta
\Gamma^{i}=\frac{1}{a}\epsilon ^{ijk}\partial_{j}\delta e_{kl}\,
dx^l\; . \ee Therefore the classical solution for the perturbed
connection is: \be\label{arealsp}
a_{ij}=\epsilon_{ikl}\partial_k\delta e_{lj}+\gamma\delta{e}'_{ij}
\ee Inserting decomposition (\ref{fourrier}) into (\ref{arealsp})
and using: \be\label{epseps}
\epsilon_{inl}k_n\epsilon^{(r)}_{lj}= -i r k \epsilon
^{(r)}_{ij} \ee we get the expression: \be\label{psia}
\Psi_a^{rp}=\gamma p \Psi'_e + r k\Psi_e \ee where, we recall,
$r=\pm$ for  $R/L$  polarizations and $p=\pm$ for graviton
positive frequency ($+$) and anti-graviton negative frequency
($-$). We have assumed that $a_{rp}=e_{rp}$, that is the
amplitudes to be promoted to creation and annihilation operators
should be equal for the metric and connection.

We can now clarify the relation between duality and helicity. They
don't align, as claimed. Inside the horizon ($k|\eta|\gg1$), equation~(\ref{ketabig})
holds and eq.~(\ref{psia}) implies that: \be
\Psi_a^{rp}=(r-ip\gamma) k\Psi_e\; . \ee Therefore the SD
connection ($\gamma=i$) is made up of the right handed positive frequency of
the graviton and the left handed negative frequency of the
anti-graviton\footnote{We note that this result is linked to the
conventions spelled out in the paragraph leading to
eq.~(\ref{sddef}). Other conventions are possible, reversing the
result in this table and the sign of $\gamma$.}. The ASD
connection contains the remaining degrees of freedom, as shown in
the table:
\begin{center}
\begin{tabular}{c | c | c}
 & $r=+$\; [R]& $r=-$ \;[L] \\
\hline
$p = +$ \; [$G$] & SD & ASD \\
$p = -$ \; [${\overline  G}$] & ASD & SD \\
\end{tabular}
\end{center}
For other values of $\gamma$ this is shared differently, and as
the modes leave the horizon ($|k\eta|\sim 1$) the classification
breaks down. Outside the horizon ($|k\eta|\ll 1$) we have \be
a_{ij}\approx \gamma Ha\delta e_{ij} \ee  Thus in this regime the
torsion free condition imposes \be\Psi_a^{rp}= \gamma p
Ha\Psi_e\ee (still assuming $a_{rp}=e_{rp}$) and therefore all
modes and polarizations appear in the connection.

\section{Reality and torsion}\label{realitytorsion}
As explained above, reality conditions should never relate
different polarizations, or modes $\vk$ and $-\vk$. If modes
propagate along the $\vk$ that labels them,
then $\vk$ and $-\vk$ modes, as well as modes with different polarizations,
are independent degrees of freedom for a real
metric, and therefore can never be constrained by reality conditions.
This is ensured by using expansions (\ref{fourrier}).
The reality of the metric ($\delta e_{ij}=\delta e_{ij}^\dagger$) then
implies:
\be
e_{r+}(\vk)=e_{r-}(\vk)
\ee
and we simply get the constraint that the graviton
and anti-graviton are identified, polarization by polarization, mode
$\vk$ by mode $\vk$. This is eminently sensible.

For the connection the situation is somewhat different. Foremost,
reality and torsion-free conditions are combined: the connection
is allowed to be complex, but only as long as it is consistent
with the metric being real, given the torsion-free condition. Thus
(for a general imaginary $\gamma$): \bea
\Re A^i&=&\Gamma^i\\
\Im A^i&=&|\gamma|\Gamma^{0i} \eea makes up the full set of
constraints. In the Hamiltonian framework one only imposes the
first of these conditions as a constraint, leaving it for the
dynamical evolution to discover the second. That is, one only
imposes the constraint: \be a_{ij}+{\overline a}_{ij}=2a \delta
\Gamma_{ij}=2 \epsilon_{ink}\partial_n \delta e_{k j} \ee which in
terms of expansion (\ref{fourrier}) becomes: \bea\label{real}
{\tilde a}_{r+} (\vk, \eta)+ {\tilde a}_{r-}(\vk,\eta) &=& 2 r k
{\tilde e}_{r+}(\vk,\eta)\\
{\tilde a}^\dagger_{r+} (\vk, \eta)+ {\tilde a}^\dagger_{r-}(\vk,\eta) &=&
2 r k {\tilde e}^\dagger_{r-}(\vk,\eta)\; ,\eea where ${\tilde a}_{rp}=
a_{rp}\Psi_a^{rp}$ and ${\tilde e}_{rp}=e_{rp}\Psi_e$. The
evolution should then imply the rest, viz: \be a_{ij}-{\overline
a}_{ij}= 2\gamma\delta e'_{ij} \ee which in terms of modes
translates into: \be\label{realim} {\tilde a}_{r+} (\vk, \eta)-
{\tilde a}_{r-}(\vk,\eta) = 2 \gamma {\tilde
e}'_{r+}(\vk,\eta)\; .\ee

\section{The Hamiltonian and Hamilton's equations}\label{secham}
We now try to rediscover the results of cosmological perturbation
theory, and those derived classically in Section~\ref{classical}, within the
Hamiltonian formalism. A proper understanding of the classical
Hamiltonian formulation is needed for quantization. As
already implied in the Introduction, the exercise is full of
surprises. Without proper care taken regarding a number of
subtleties, one actually does {\it not} recover the results of
standard cosmological perturbation theory.

For a general Immirzi parameter the gravitational Hamiltonian is
given by: \bea {\cal H}&=&\frac{1}{2l_P^2}\int d^3x N E^a_i E^b_j
\Big[\epsilon_{ijk}(F^k_{ab}+H^2 \epsilon _{abc} E^c_k)\nn
&&-2(1+\gamma^2)K^i_{[a} K^j_{b]}\Big] \label{fullH}\eea where \be\label{extK}
K^i_a=\frac{A^i_a-\Gamma^i_a(E)}{\gamma} \ee is the extrinsic
curvature of the spatial surfaces (on shell this becomes
$K^i_a\approx \Gamma^{0i}_a$, something that should be discovered
by Hamilton's equations). To this volume integral one must add a
boundary term~\cite{gibbonshawking,teitelboim,btashtekar}:
\be\label{boundary} {\cal H}_{BT}=-\frac{1}{l_P^2}\int d\Sigma_a
N\epsilon_{ijk} E^a_i E^b_j A_{bk}\; . \ee This term may be zero
(e.g. if the manifold has no boundary), but
otherwise ignore it at your peril. Its vanishing is
often ensured by imposing suitable fall-off
conditions~\cite{gravitons,btashtekar}, but these are
blatantly violated by plane waves. Therefore in the study of mode
solutions (and their Hamiltonian) it is essential to include the
boundary term (this matter is usually swept under the carpet by
performing an ad-hoc integration by parts in order to obtain the
``right result'').

Strictly speaking, (\ref{fullH}) constitutes the Hamiltonian
constraint;  the full Hamiltonian
is made up of two other constraints, the Gauss constraint:
\be\label{gauss} G_i=D_a E^a_i=\partial_a E^a_i + \epsilon_{ijk}A^j_a E^a_k\approx 0 \ee and the vector
constraint \be\label{vec} V_b=E^a_iF^i_{ab}\approx0 \ee (a
combination produces the so-called diffeomorphism constraint).
These are automatically satisfied to first order by the mode
decompositions chosen. A comment on the impact of this for
the commutation relations will be made in Section~\ref{coms}.

The dynamics is specified by the Hamiltonian together with
sympletic structure \be
\label{PBnonpert}\{A^i_a(\vx),E^b_j(\vy)\}=\gamma
l_P^2\delta^b_a\delta^i_j\delta(\vx-\vy)\; . \ee We'll now rediscover
the usual equations of cosmological perturbation theory in this
framework. For clarity we will make our various points restricting
ourselves to $\gamma=\pm i$ (as we did in~\cite{prl}). In the last
subsection we'll then list the corresponding results for
the general case.

\subsection{Hamilton's equations for $\gamma=\pm i$}
The full Hamilton equations for $\gamma=\pm i$ are: \bea
{A^{i}_a}'&=&\{A^i_a,{\cal H}\}=\gamma N\epsilon_{ijk}
E^b_j{\left(F^k_{ab}+\frac{3}{2}H^2\epsilon_{abc}E^c_k\right)}\nonumber\\
{E^{a}_i}'&=&\{E^a_i,{\cal H}\}=-\gamma\epsilon_{ijk}D_b(N E^a_j
E^b_k)\; . \eea It's easy to check that the background solution
(\ref{zeroth1})-(\ref{zeroth}) solves the Hamiltonian constraint
${\cal H}\approx 0$, as well as Hamilton's equations. In
performing this exercise note that using conformal time, the lapse
{\it density} is $N=1/a^2$. This is a simple test on consistency
of conventions. The background solution also trivially satisfies
the Gauss and vector constraints (\ref{gauss})-(\ref{vec}).

If we perturb these equations via (\ref{pertE}) and (\ref{pertA})
we find: \bea a'_{ij}&=&2
\gamma H^2 a^2\delta e_{ij}-\gamma \epsilon_{inm}
\partial_n a_{mj}\label{ham1}\\
\delta e'_{ij}&=&-\gamma(a_{ij}-\epsilon_{inm}\partial_n\delta
e_{mj})\; . \label{ham2}\eea  The Hamilton equation for $\delta
e_{ij}$ is simply (\ref{arealsp}) for $\gamma=\pm i$, i.e. a
statement that on-shell the connection is the torsion-free SD or
ASD connection (or alternatively, a confirmation that the
extrinsic curvature is $\delta K_{ij}=e'_{ij}$; cf.
Eqn~(\ref{extK})). Combining equations (\ref{ham1}) and (\ref{ham2}) we obtain the familiar second order
equation for $\delta e_{ij}$: \be \delta
e''_{ij}-{\left(\partial^2 +\frac{2}{\eta^2}\right)}\delta
e_{ij}=0 \ee equivalent to (\ref{veq}) in Appendix I. This shows
that classically, the standard formalism of cosmological
perturbation theory and the Hamiltonian Ashtekar framework are
equivalent.

\subsection{The perturbative status of the Hamiltonian
constraint}\label{sechamconst} As we've seen it's easy to find the
perturbation equations by perturbing the full Hamilton's
equations. However, locating the perturbed Hamiltonian within the
full theory is more subtle. This is to be expected from the fact
that the usual Hamiltonian constraint ${\cal H}\approx 0$ cannot
apply to the perturbative Hamiltonian, since gravitons do have
dynamics.

The first order Hamiltonian is trivially zero (once the other
constraints are used). The second order Hamiltonian is made of two
terms: \be ^2{\cal H}={^{2}_1{\cal H}}+{^{2}_2{\cal H}}\, , \ee
where ${^{2}_1{\cal H}}$ contains products of
first order perturbations and ${^{2}_2{\cal H}}$ contains second order
perturbations in the triad and connection. Only the total must
vanish on shell. The first term provides a candidate for the
Hamiltonian to be identified with that of the second quantized QFT.
The second contains the backreaction or compensation
resulting from the non-linearity of the gravitational field,
ensuring that the Hamiltonian constraint is satisfied, whilst a
strictly positive ${^{2}_1{\cal H}}$ remains possible. We'll
concentrate on term ${^{2}_1{\cal H}}$ and
so ignore the Hamiltonian constraint
to second order (which would only provide us with information on
the backreaction term).

Before we proceed with the algebra we need to stress a crucial point: In
the formulation we are using, {\it the Hamiltonian is not real}.
Of course, after the constraints are imposed the Hamiltonian
becomes weakly zero and therefore real. However, off-shell we have
to deal with an intrinsically complex Hamiltonian. Furthermore
this distinctive feature propagates into perturbation theory. The
Hamiltonian constraint doesn't apply to the piece of the
Hamiltonian, $^2_1{\cal H}$, which provides the raw material for
the second quantized theory living inside the full
non-perturbative theory. Therefore $^2_1{\cal H}$ is complex, and
we have to live with it. Some of the novelties to be derived in
this paper trace their origin directly to this fact. As we will
explain later, however, the Hamiltonian is still Hermitian with respect
to an inner product still to be defined (and the two matters,
complexity and Hermiticity, should not be confused).

By expanding the Hamiltonian to second order we find: \bea
\label{hampert} {^{2}_1{\cal H} }&=&\frac{1}{2l_P^2}\int d^3x [
-a_{ij}a_{ij} +2 \epsilon_{ijk}
\delta e_{li}\partial_j a_{kl} \nonumber\\
&& -2 \gamma Ha \delta e_{ij}a_{ij} -2H^2 a^2 \delta e_{ij} \delta
e_{ij} ]\;. \eea Note that in obtaining (\ref{fullH}) from the
usual ADM action (by means of an extension of the phase space and
a canonical transformation), a number of algebraic manipulations
are needed in which the Gauss (or ``rotational'') constraint and
the torsion free condition are used (see~\cite{thbook}; chapters
1 and 4). However, when evaluating the Hamiltonian to second
order, ${}^2_1{\cal H}$, we have to bear in mind that 
expansions (\ref{fourrier}) only solve the Gauss constraint 
to linear order, so that to second order \be
{}^2_1G_i=-\epsilon_{ijk}a^j_a \delta e^a_k\neq 0\; . \ee Thus,
terms linear in ${}^2_1G_i$, multiplied by zero order variables,
will appear in ${}^2_1{\cal H}$. Likewise, with solution
(\ref{deltagamma}), the torsion $T^a= d e^a +\Gamma^a_b \wedge
e^b$ only vanishes to first order, and has a non-zero contribution
quadratic in first order perturbation variables, i.e.
${}^2_1T^a\neq 0$. This affects several manipulations leading to
${}^2_1{\cal H}$, and even the definition of the covariant
derivative with respect to the Ashtekar connection (note that the
Gauss constraint is initially a ``rotational'' constraint,
$G_i=\epsilon _{ijk}K_a^jE^a_k$, and not a proper Gauss law;
see~\cite{thbook}, pp.124). It can be checked that all extra terms
in ${}^2_1{\cal H}$ resulting from these considerations form a
full divergence irrelevant for the purpose of this paper.

\subsection{Two further subtleties}\label{twosubtle}
It may seem that we have identified the portion of the full
Hamiltonian to be associated with the perturbative Hamiltonian to
second order, but this is not the case. It's easy to check that
(\ref{hampert}) does not reduce to the expected expression
(\ref{cosmoH}) in Appendix I on-shell, i.e. using (\ref{arealsp}).
This is due to two reasons.

Firstly, one must add the corresponding boundary term
(\ref{boundary}) at the same order and level in perturbation
theory (i.e. second order terms quadratic in first order
variables). This is: \be {^{2}_1{\cal H}
}_{BT}=\frac{1}{l_P^2}\int d\Sigma_i \epsilon_{ijk}\delta e_{lj}
a_{lk} \ee which should be brought down to the interior of the
region in the form of a divergence. When this is done we obtain:
\bea\label{effectham0} {\cal H} _{eff}&=&\frac{1}{2l_P^2}\int d^3x
[ -a_{ij}a_{ij} -2 \epsilon_{ijk}
(\partial_j\delta e_{li}) a_{kl} \nonumber\\
&& -2 \gamma Ha \delta e_{ij}a_{ij} -2H^2 a^2 \delta e_{ij} \delta
e_{ij} ]\; \eea and if we can ignore terms in $H$  it's easy to
check that classically (on-shell)
this is nothing but the usual expression for the stress-energy tensor
of gravitational waves, with a kinetic and a gradient term as
usual. However, if terms in $H$ cannot be neglected this is still
{\it not} Eq.~(\ref{cosmoH}).

In order to understand why, it's well worth looking at
``perturbative'' expressions \be A^i_a=\gamma Ha \delta^i_a +
\frac{a^i_a}{a}\label{eexp}\ee\be E^a_i=a^2\delta^a_i -a \delta
e^a_i \label{aexp}\ee in a different way: {\it they represent a
canonical transformation}. Above all we have traded canonical
variables $(A^i_a,E^b_j)$ with variables $(a^i_a,\delta e^b_j)$;
and only then, in what might be properly called perturbation
theory, have we assumed the latter to be small, so that a
truncation scheme can be set up. However the transformation can
always be carried out, even when $(a^i_a,\delta e^b_j)$ are not
small and no truncations are applied.

This way of thinking has several advantages. Firstly it
permits a rigorous derivation of the sympletic structure for the
new variables: \be\label{PBperts} \{a^i_a(\vx),\delta e^b_j(\vy)\}
= -\gamma l_P^2\delta^b_a\delta^i_j\delta(\vx-\vy)\;  \ee (where
the minus sign appears because $\delta e_{ij}$ is the perturbation
in the triad, not the densitized inverse triad)~\footnote{According to some
conventions we should endow the new variables $(a^i_a,\delta
e^b_j)$ with the same brackets as $(A^i_a,E^b_j)$. A minus sign
would then appear in front of the perturbative Hamiltonian.}. We'll use these
Poisson brackets to define the quantum theory in the next Section.
Contrary to a common myth we are not ``freezing'' the
background, and allowing the perturbations to ``quantum
fluctuate''\footnote{A related myth is that there is no Poisson
bracket between zero and first order variables because they ``fluctuate
independently''.}. We are simply replacing a canonical pair by another,
which is particularly suited to our problem, since
its classical variables can be assumed  to be small.

Furthermore the matter is far from pedantic if $a$ is not a
constant, because the transformation is then time dependent.
Therefore the new Hamiltonian (sometimes represented by $K$;
see~\cite{goldstein}) is not simply the old Hamiltonian written in
terms of the perturbations $(a^i_a,\delta e^b_j)$ (and then
possibly truncated). Instead it must be replaced by: \be {\cal
K}={\cal H}+\frac{\partial F}{\partial \eta} \ee where $F$ is the
generating function of the canonical transformation, if we want to
obtain an equivalent pair of Hamilton's equations. It can be
easily checked that if we take the perturbed (old) Hamiltonian
(\ref{effectham0}) and work out Hamilton's equations following
from (\ref{PBperts})  we'd be presented with a result inconsistent
with (\ref{ham1}) and (\ref{ham2}), obtained by evaluating
Hamilton's equations and then perturbing. This is prevented by
using \be \frac{\partial F}{\partial \eta}=
\frac{\gamma}{l_P^2}\int d^3x  Ha \delta e_{ij}a_{ij}\; .\ee In
Appendix II this generating functional is derived for general
values of $\gamma$.

Therefore, we obtain as the Hamiltonian for the new variables:
\bea\label{effectham} {\cal H} _{eff}&=&\frac{1}{2l_P^2}\int d^3x
[ -a_{ij}a_{ij} -2 \epsilon_{ijk}
(\partial_j\delta e_{li}) a_{kl} \nonumber\\
&& -2H^2 a^2 \delta e_{ij} \delta e_{ij} ]\; \eea This Hamiltonian
should be identified with the Hamiltonian of the second quantized,
effective quantum field theory representing the theory
perturbatively. And indeed, ``on-shell'', i.e. using
(\ref{arealsp}), this does finally reduce to  Eq.~(\ref{cosmoH}).

\subsection{A general $\gamma$}
The various points made in the previous subsections, for
$\gamma=\pm i$, remain valid in the general case, but the algebra is far
more complex. Here we list the corresponding results.
Hamilton's
equations for a general $\gamma$ take the form: \bea
{A^{i}_a}'&=&\gamma N\epsilon_{ijk}
E^b_j{\left(F^k_{ab}+\frac{3}{2}H^2\epsilon_{abc}E^c_k\right)}\nonumber\\
&-&\gamma(1+\gamma^2)NE^b_j (K^j_b K^i_a - K^j_a K^i_b)\nonumber\\
&-&\frac{1+\gamma^2}{l_P^2}\int d^3y N E^b_j E^c_k \{A^i_a (x),
\Gamma^j_{[b} \Gamma^k_{c]} \}\\
{E^{a}_i}'&=&-\gamma\epsilon_{ijk}D_b(N E^a_j
E^b_k)\nonumber\\
&&+(1+\gamma^2)N(E^a_iE^b_j - E^a_jE^b_i)K^j_b \; . \eea As far as
we are aware there is no simple expression for the Poisson bracket
$\{A,\Gamma (E)\}$, and so we left the last term of the first
equation unexpanded~\footnote{In a number of treatments $\Gamma$
is seen as an independent variable, which only becomes $\Gamma(E)$
on shell, and which thus commutes with $A$. This doesn't clear up
the messy last term in the Hamilton equation for $A$, should we
need an explicit expression.}. It is again easy to prove that the
zero order solution (\ref{zeroth1})-(\ref{zeroth}) satisfies these
equations as well as all the constraints. It is also
straightforward to infer a complete closed form expression for the
Hamilton equations for the perturbations: \bea a'_{ij}&=&2 \gamma
H^2 a^2\delta e_{ij}-\gamma \epsilon_{inm}
\partial_n
a_{mj}\nonumber\\
&+&\frac{1+\gamma^2}{\gamma}\epsilon_{inm}\partial_n
(a_{mj}-\epsilon_{mkl}\partial_k\delta e_{lj})\\
\delta
e'_{ij}&=&\frac{1}{\gamma}(a_{ij}-\epsilon_{inm}\partial_n\delta
e_{mj})\; . \eea Again the Hamilton equation for $\delta e_{ij}$ is
equivalent to (\ref{arealsp}), and so a statement that the
connection is torsion free. Combined with the Hamilton equation
for $a_{ij}$ we obtain a second order equation for $\delta e_{ij}$
from which $\gamma$ drops out, as it should. This is obviously
(\ref{veq}) again, proving consistency with cosmological
perturbation theory for all $\gamma$.

By expanding the Hamiltonian to second order in first order
variables we  obtain the counterpart of (\ref{hampert}):
\bea\label{totalham0} &&{}^2_1{\cal H}=\frac{1}{2l_P^2}\int
d^3x\biggl[\frac{1}{\gamma^2}a_{ij}a_{ij} +2 \epsilon_{ijk} \delta
e_{li}\partial_j a_{kl}
\nonumber \\
&&+\frac{2}{\gamma} Ha \delta e_{ij}a_{ij}-
2\frac{1+\gamma^2}{\gamma} Ha \delta e_{ij} \epsilon_{ikl}
(\partial_k \delta e_{lj})
\nonumber \\
&&- \frac{1+\gamma^2}{\gamma^2}\left[\epsilon_{ikl}(\partial_k
\delta e_{lj})a_{ij} + \epsilon_{ikl}a_{ij}(\partial_k \delta
e_{lj})\right]
\nonumber\\
&& +
\frac{1+\gamma^2}{\gamma^2}\epsilon_{ikl}\epsilon_{jmn}(\partial_k
\delta e_{lj})(\partial_m \delta e_{ni}) - 2H^2a^2\delta e_{ij}
\delta e_{ij} \biggr]\, .\nonumber  \eea
where we have been careful with the ordering, with an eye on
quantization.
As before the boundary term modifies the second term in
the first line. The second line is cancelled by the extra term
associated with the canonical transformation:  \bea\label{genfun} \frac{\partial
F}{\partial \eta}= -\frac{1}{\gamma l_P^2}\int d^3x Ha \delta
e_{ij}\left[a_{ij}-(1+\gamma^2)\epsilon_{inm}\partial_n\delta
e_{mj}\right]\; .\nonumber  \eea This can be inferred by examining
the behavior of Hamilton's equations under time-dependent
rescalings of $a_{ij}$ and $\delta e_{ij}$. If we try to combine them into a single second order
equation, the term proportional to $(\gamma^2+1)$
is obviously necessary: This is because $K=(A-\Gamma)/\gamma$ contains
two terms which scale differently. So in order for the time derivatives to scale
in a form producing consistent equations, we need to add the
second term. With these considerations the effective Hamiltonian,
to be used in the second quantized theory, is therefore:
\bea\label{totalham}
&&{\cal H}_{eff}=\frac{1}{2l_P^2}\int d^3x\biggl[\frac{1}{\gamma^2}a_{ij}a_{ij} - 2H^2a^2\delta e_{ij} \delta e_{ij} \nonumber \\
&&+ \left(1-\frac{1}{\gamma^2}\right)\epsilon_{ikl}(\partial_k
\delta e_{lj})a_{ij} -
\left(1+\frac{1}{\gamma^2}\right)\epsilon_{ikl}a_{ij}(\partial_k
\delta e_{lj})
\nonumber\\
&& +
\left(1+\frac{1}{\gamma^2}\right)\epsilon_{ikl}\epsilon_{jmn}(\partial_k
\delta e_{lj})(\partial_m \delta e_{ni}) \biggr]\, . \eea

For completeness we present the perturbative equations for the
theory in terms of $\delta K_{ij}$ and $\delta e_{ij}$, i.e. in
the extended ADM form, before a canonical transformation is
applied~\cite{thbook}. This corresponds to the Palatini-Kibble
limit, $|\gamma|\rightarrow \infty$, with a $SU(2)$ extension of
the phase space. The equations are:\bea \delta K_{ij}'&=&2H^2a^2
\delta e_{ij} -\epsilon
_{inm}\partial_n\delta\Gamma_{mj}\\
\delta e_{ij}'&=& K_{ij}\; . \eea Undoubtedly these are the
simplest equations, together with $\gamma=\pm i$. It is very easy
to check that these equations can be combined into an equation of the
form of (\ref{veq}).

\section{Commutation relations in terms of modes}\label{coms}
In order to quantize the theory we need to replace Poisson
brackets (\ref{PBnonpert}) and (\ref{PBperts}) by
commutators~\footnote{If the reality conditions are to be seen as
second class constraints at the quantum level, we should identify
the Dirac brackets  at this step. We shall examine this
possibility in a future article.}.
Thus, we obtain  equal-time commutation relations:
\be\label{unfixedcrs} \left[A^i_a(\vx),E^b_j(\vy)\right] = i\gamma
l_P^2\delta^b_a\delta^i_j\delta(\vx-\vy)\; , \ee
and: \be\label{unfixedcrs1} \left[a^i_a(\vx),\delta
e^b_j(\vy)\right] = -i\gamma
l_P^2\delta^b_a\delta^i_j\delta(\vx-\vy)\; . \ee
These are the
commutation relations before the Gauss constraint is enforced.
They must be replaced by a suitably TT projected $\delta$-function
after gauge fixing.

Dropping the indices for the moment, we can separate positive and
negative frequencies in equation (\ref{fourrier}) as  $\delta e= \delta e^
++\delta e^-$ where: \bea \delta e^+(\vx,\eta)&=&\int \frac{d^3
k}{(2\pi)^{\frac{3}{2}}} e^+(\vk,\eta)
e^{i\vk\cdot \vx}\\
\delta e^-(\vx,\eta)&=&\int \frac{d^3 k}{(2\pi)^{\frac{3}{2}}}
e^{-\dagger}(\vk,\eta) e^{-i\vk\cdot \vx} \eea and likewise for
$a=a^++a^-$. The only non-vanishing equal-time commutators must
be: \be [a^+(\vx),\delta e^-(\vy)]= [a^-(\vx),\delta e^+(\vy)]=
-i\gamma \frac{l_P^2}{2} \delta(\vx-\vy)\; , \ee so that
\be\label{aecrs} [a^+(\vk),e^{-\dagger}(\vk ')]= -
[a^-(\vk),e^{+\dagger}(\vk ')]= -i\gamma
\frac{l_P^2}{2}\delta(\vk-\vk ')\; , \ee where the minus sign in
the second commutator appears because $\gamma$ is imaginary. We
stress that with our conventions and boundary conditions $\vk$ and
$-\vk$ modes propagate in different directions and so they are
independent. Therefore their amplitudes must commute and
(\ref{aecrs}) had to be proportional to $\delta(\vk-\vk ')$. This
is to be contrasted with some of the literature.

Upon gauge fixing these expressions take the specific
forms~(\ref{fourrier}). This results in the TT-fixed commutators:
\be\label{fixedcrs} [{\tilde a}_{rp}(\vk),{\tilde
e}_{sq}^\dagger(\vk ')] =-i\gamma p
\frac{l_P^2}{2}\delta_{rs}\delta_{p{\bar q}} \delta(\vk-\vk ')\; ,
\ee
(where ${\overline q}=-q$). With these relations we get the
expected~\cite{weinberg} version of (\ref{unfixedcrs1}):
\be\label{gaugefixed} [a_{ij}(\vx),\delta e_{kl}(\vy)]=-i\gamma
l_P^2P_{ijkl}(\vx-\vy)\; , \ee with \be P_{ijkl}(\vx)=\int
\frac{d^3 k}{(2\pi)^{3}}\sum_r \epsilon^{r} _{ij}(\vk)\epsilon
^{r\star}_{kl} (\vk) e^{i\vk\cdot\vx} \;  \ee (as before we mix
$i$ and $a$ indices, raising and lowering them with a Kroenecker
delta).  Using the polarization completeness relations this can
also be written as~\cite{weinberg}: \be P_{ijkl} =\int \frac{d^3
k}{(2\pi)^{3}} \Pi_{ijkl}(\vk) e^{i\vk\cdot (\vx-\vy)} \; , \ee
with \bea \Pi_{ijkl}(\vk)&=&\frac{1}{2}[ (\delta_{ik}-{\hat
k}_i{\hat k}_k)(\delta_{jl}-{\hat k}_j{\hat k}_l)
\nonumber\\
&&+(\delta_{il}-{\hat k}_i{\hat k}_l)(\delta_{jk}-{\hat k}_j{\hat
k}_k)
\nonumber\\
&&-(\delta_{ij}-{\hat k}_i{\hat k}_j)(\delta_{kl}-{\hat k}_k{\hat
k}_l)] \eea or equivalently \bea \Pi_{ijkl}(\vx)&=&\frac{1}{2}[
(\delta_{ik}-\frac{\partial_i\partial_k}{\partial^2})(\delta_{jl}-\frac{\partial_j\partial_l}{\partial^2})
\nonumber\\
&&+(\delta_{il}-\frac{\partial_i\partial_l}{\partial^2})(\delta_{jk}-\frac{\partial_j\partial_k}{\partial^2})
\nonumber\\
&&-(\delta_{ij}-\frac{\partial_i\partial_j}{\partial^2})(\delta_{kl}-\frac{\partial_k\partial_l}{\partial^2})]\delta(\vx)\;
. \eea We point out that (\ref{fixedcrs}) isn't exactly what could
have been guessed, say, by quantizing a complex scalar field. The
reason is that the action and Hamiltonian are not real before the
reality conditions are imposed (not of the form
$\partial_\mu\phi\partial^\mu\phi^\star$). Thus the classical
relation between the variable $A$ and its conjugate $E$ does not
involve complex conjugation (unlike in $\Pi=\dot\phi^\dagger$). As
a result we get a $\delta_{p{\overline q}}$ in the result, i.e.
non-vanishing commutators involve the negative frequencies of one
variable and the positive frequencies of the other.  This wouldn't
happen if the Hamiltonian were manifestly real. Another oddity is
the $p$ factor in the commutator: it is present only because
$\gamma$ has been assumed to be imaginary.

\section{The quantum Hamiltonian and a possible representation for the inner product}
\label{quantum ham} We now expand the Hamiltonian found in the
previous section into Fourier modes, thereby identifying the
combinations of metric and connection variables to be equated with
the graviton. This is a non-trivial exercise (particularly for
$\gamma\neq \pm i$) and we invariably find twice as many particles as expected.
This is because the reality conditions are yet to be imposed. At
the quantum level, this is done via the choice of inner product
with which the Hilbert space is endowed. Two out of the four modes
are then seen to be unphysical and can be removed from the Hilbert space.
These spurious modes, as it turns out, have negative energy and don't exist classically (i.e.
are zero on-shell).

\subsection{Inside the horizon, with $\gamma=\pm i$}
We consider the limit $k|\eta|\gg 1$, i.e. modes inside the
horizon. We assume an $EEF$ ordering but what follows can be
adapted to other orderings, with the general result explained in
Section~\ref{order}. Inserting expansions (\ref{fourrier}) into
(\ref{effectham}) we find: \bea\label{hameff} {\cal
H}_{eff}&=&\frac{1}{l_P^2}\int d^3k\sum_r
g_{r-}(\vk)g_{r+}(-\vk)+g_{r-}(\vk)g_{r-}^\dagger(\vk)\nonumber\\
&+&g_{r+}^\dagger(\vk)g_{r+}(\vk)+
g_{r+}^\dagger(\vk)g_{r-}^\dagger(-\vk)\; , \eea with: \bea\label{gmodes}
g_{r+}(\vk)&=&{\tilde a}_{r+}(\vk)\\
g^\dagger_{r+}(\vk)&=&-{\tilde a}^\dagger_{r-}(\vk) +2kr {\tilde e}_{r-}^\dagger(\vk)\\
g_{r-}(\vk)&=&-{\tilde a}_{r+}(\vk) +2kr {\tilde e}_{r+}(\vk)\\
g_{r-}^\dagger(\vk)&=&{\tilde a}^\dagger_{r-}(\vk)\label{gmodes1} \eea where we
used
$\epsilon^r_{ij}(\vk)\epsilon^{s\star}_{ij}(\vk)=2\delta^{rs}$
(note that with our conventions
$\epsilon_{ij}^r(-\vk)=\epsilon^{r\star}_{ij} (\vk)$). We have
identified (anti)-graviton creation and annihilation operators,
$g_{rp}^\dagger$ and $g_{rp}$, as in~\cite{gravitons}. From
(\ref{fixedcrs}) they inherit the algebra: \be\label{galgebra}
[g_{rp}(\vk),g^\dagger_{sq}(\vk')]=-i\gamma l_P^2 (pr)k
\delta_{rs}\delta_{pq} \delta(\vk-\vk')\; . \ee This Hamiltonian
has several strange features. For a given $\vk$ (and here this
means modes moving along $\vk$ and not $-\vk$) we find 4, not 2
independent modes ($r=\pm 1$ and $p=\pm$ 1). As in Witten's
argument~\cite{wittenym}, half of these particles have negative
energy (those with $i\gamma= pr$). For example, for the SD
connection the left ``graviton'' ($r=-1$ and $p=1$) and the right
``anti-graviton'' ($r=1$ and $p=-1$) carry negative energy. The
Hamiltonian also contains pathological particle production terms:
the first and last terms in (\ref{hameff}). Such terms, coupling
$\vk$ and $-\vk$ modes, are pump terms \cite{grishchuk}
representing pair production, and must be unphysical in a
Minkowski brackground or for $k|\eta|\gg 1$.

Both of these pathological features are not present for classical
solutions, i.e. they vanish on-shell (by imposing the solutions
found in Section~\ref{classical}). For example, for $\gamma=i$ we
have $a_{R-}\approx 0$ and $a_{L+}\approx 0$. Thus the negative
energy modes don't exist classically and the pump terms are
identically zero. Quantum mechanically these features must be
erased by removing the spurious modes present in the full Hilbert
space of the second quantized theory. This is done by an
appropriate choice of inner product, representing the reality
conditions at the quantum level, as we now show.

Notice first that the reality conditions amount to demanding that
$g_{rp}^\dagger$ be indeed the hermitian conjugate of $g_{rp}$.
This fully fixes the inner product~\cite{tate,pulbook,gravitons}.
We work in a holomorphic representation for wavefunctions $\Phi$
which diagonalizes $g^\dagger_{rp}$, i.e.: \be
g^\dagger_{rp}\Phi(z)= z_{rp}\Phi(z)\ee ($z$ represents
collectively all the $z_{rp}(\vk)$). Then, (\ref{galgebra})
implies: \be \label{grop} g_{rp}\Phi =-i\gamma l_P^2 (pr)k
\frac{\partial \Phi}{\partial z_{rp}} \; .\ee
With the Ansatz for the inner product: \be\label{ansatz}{\langle
\Phi_1 | \Phi_2\rangle}=\int d z d {\bar {z}} e^{\mu(z,{\bar z})}
{\bar \Phi_1}({\bar z}) \Phi_2 (z)\ee the formal condition
${\langle \Phi_1 |g_{rp}^\dagger |\Phi_2\rangle}={\overline
{\langle \Phi_2 | g_{rp} |\Phi_1\rangle}}$ therefore requires: \be
\mu(z,{\bar z})=\int d{\vk}\sum_{rp}\frac{pr}{ik\gamma l_P^2}
z_{rp}(\vk){\bar z} _{rp}(\vk)\; , \ee fixing ${\langle \Phi_1 |
\Phi_2\rangle}$. Integrating $g_{rp}\Phi_0=0$ leads, in this
representation, to the vacuum \be \Phi_0={\langle z|0\rangle}=1\;
.\ee Particle states are monomials in the respective variables,
\be\Phi_n={\langle z|n\rangle}\propto (g_{rp}^\dagger)^n \Psi_0=
z_{rp}^n.\ee  Thus, with the inner product just derived these
aren't normalizable for $i\gamma= pr$. Therefore such modes should
be excluded from the physical Hilbert space, and this removes all
pathologies found in the Hamiltonian. We stress that the quantum
modes we have disqualified don't exist classically (see discussion
after (\ref{psia})). For example for $\gamma=i$ the only physical
modes are $g_R^{ph}=g_{R+}$ and $g^{ph}_L=g_{L-}$.

We therefore regain the usual physical Hamiltonian, but with one
major difference. For $\gamma=i$ we get \be{\cal
H}^{ph}_{eff}\approx \frac{1}{l_P^2}\int d\vk \,  (g^{ph}_L
{g^{ph}_{L}}^{\dagger} + {{g}_{R}^{ph}}^\dagger {g}^{ph}_{R}) \ee
and so only the left handed graviton needs to be normal ordered
and produces a vacuum energy. For the ASD connection only the
right handed graviton produces vacuum energy. This chirality
feature traces directly to the fact that the Hamiltonian is not
real, a priori, as explained before.

\subsection{Inside the horizon, with $\gamma^2\neq -1$}
We now examine the more general case of a connection which isn't
SD or ASD ($\gamma=\pm i$). We shall only consider imaginary
$\gamma$ leaving the case of a generally complex $\gamma$ for a
further publication~\cite{future}. The exercise is straightforward
but the algebra is cumbersome.

The Hamiltonian is now quite intricate (see Appendix III), making it difficult
to identify the graviton modes. Instead it's easier
to look at the classical solution (obtained from the Lagrangian
formalism; see Section~\ref{classical}) and infer combinations of
${\tilde a}_{rp}$, ${\tilde e}_{rp}$ and their conjugates equivalent to the $g_{rp}$
and $g^\dagger_{rp}$ listed in (\ref{gmodes}) to (\ref{gmodes1}). One of these
combinations should be zero on-shell, and represent the unphysical
quantum mode. The other should commute with it, given commutation
relations (\ref{fixedcrs}) or (\ref{galgebra}), for all
values of $\gamma$. These requirements suffice to determine: \bea
G_{r{\cal P_+}}&=&\frac{(r-i\gamma)g_{r+}-(r+i\gamma)g_{r-}}{-2\gamma i}\\
G_{r{\cal P_-}}&=&\frac{(r+i\gamma)g_{r+}-(r-i\gamma)g_{r-}}{-2\gamma i}\, .
\eea
We have introduced a new index ${\cal P}={\cal P_+},{\cal P_-}$ to
label physical and non-physical modes. The notation may look strange
but it has the virtue of avoiding confusions with $p=\pm$ used for
positive and negative frequency. Except for the cases of $\gamma=\pm i$
the two don't align. Thus,
${\cal P}={\cal P_+}=1$ denotes the physical modes, which shouldn't vanish
classically, and quantum mechanically are expected to have
positive energy and norm; ${\cal P}={\cal P_-}=-1$ denotes
modes that classically
vanish, and quantum mechanically are expected to have negative
energy and norm.

It is easy to see that on-shell and enforcing the
reality conditions,
$G_{r{\cal P_-}}\approx 0$ and $G_{r{\cal P_+}}\approx 2rk e_r$,
so our first requirement is satisfied. Furthermore we find
commutation relations: \bea\label{Galgebra}
\left[G_{r{\cal P}}(\vk),G^{\dagger}_{s{\cal P}}(\vk')\right]&=&{\cal P} k l_P^2
\delta_{rs}
\delta(\vk-\vk')\\
\left[G_{r{\cal P_+}}(\vk),G^{\dagger}_{s{\cal P_-}}(\vk')\right]&=&0 \eea as
required. Notice that these combinations  are {\it not} a rotation
upon the original modes found for $\gamma=\pm i$. A pure rotation
would leave $[G_{r{\cal P_+}},G^{\dagger}_{r{\cal P_-}}]\neq 0$.

It is useful to write these modes directly in terms of triad and
connection variables. Their general (off-shell, before imposing reality
conditions) expression is: \be \label{Gr+}G_{r{\cal P_+}} =
\frac{-r}{i\gamma}{\left({\tilde a}_{r+}- k(r+i\gamma){\tilde
e}_{r+}\right)}\ee\be
\label{Gr+dagger}G^{\dagger}_{r{\cal P_+}}=\frac{r}{i\gamma}({\tilde
a}^\dagger_{r-}- k(r-i\gamma){\tilde e}^\dagger_{r-})\ee\be
G_{r{\cal P_-}}=\frac{-r}{i\gamma}({\tilde a}_{r+}- k(r-i\gamma){\tilde
e}_{r+})\ee
\be \label{Gr-dagger}G^{\dagger}_{r{\cal P_-}}=\frac{r}{i\gamma}({\tilde
a}^\dagger_{r-}- k(r+i\gamma){\tilde e}^\dagger_{r-})\; . \ee It
is a straightforward algebraic exercise to show that inside the horizon
(i.e. setting $H=0$) the off-shell Hamiltonian of Appendix III can
be written as:
\begin{widetext}
\bea
&&{\cal H}_{eff}=\frac{1}{2 l_P^2}\int d^3k\sum_r
-(1+i\gamma r)G_{r{\cal P_+}}(\vk)G_{r{\cal P_-}}(-\vk)
-(1-i\gamma r)
G_{r{\cal P_-}}(\vk)G_{r{\cal P_+}}(-\vk)\nonumber\\
&&+(1+i\gamma r)G_{r{\cal P_+}}(\vk)G_{r{\cal P_+}}^\dagger(\vk)+(1-i\gamma r)
G^\dagger_{r{\cal P_+}}(\vk)G_{r{\cal P_+}}(\vk)
+(1-i\gamma r)G_{r{\cal P_-}}(\vk)G_{r{\cal P_-}}^\dagger(\vk)+(1+i\gamma r)
G^\dagger_{r{\cal P_-}}(\vk)G_{r{\cal P_-}}(\vk)\nonumber\\
&&-(1-i\gamma r)G^\dagger_{r{\cal P_+}}(\vk)G^\dagger_{r{\cal P_-}}(-\vk)
-(1+i\gamma
r) G^\dagger_{r{\cal P_-}}(\vk)G^\dagger_{r{\cal P_+}}(-\vk)\eea
\end{widetext}
(In spite of its horrendous appearance, this is nothing but a
generalization of Eqn.~(\ref{hameff})). The Hamiltonian contains the
same pathologies previously found for $\gamma=\pm i$, and again
these are removed once the reality conditions are taken into
account. This can be done with the choice of inner product.

As before, we work in a holomorphic representation which
diagonalizes $G^\dagger_{r{\cal P}}$, i.e.: \be G^\dagger_{r{\cal P}}\Phi(z)=
z_{r{\cal P}}\Phi(z)\ee Then, (\ref{Galgebra}) implies: \be \label{Grop}
G_{r{\cal P}}\Phi = {\cal P}k l_P^2 \frac{\partial \Phi}{\partial z_{r{\cal P}}} \;
.\ee
Formally nothing seems very different, but note that the variables
$z_{r{\cal P}}$ expressed in terms of metric and connection variables are
different from the $\gamma=\pm i$ case. With the same Ansatz
(\ref{ansatz}) and the same formal condition ${\langle \Phi_1
|G_{r{\cal P}}^\dagger |\Phi_2\rangle}={\overline {\langle \Phi_2 |
G_{r{\cal P}} |\Phi_1\rangle}}$ we obtain: \be \mu(z,{\bar
z})=\int d{\vk}\sum_{r{\cal P}}\frac{-{\cal P}}{k l_P^2}z_{r{\cal P}}(\vk){\bar z}
_{r{\cal P}}(\vk)\; , \ee fixing the inner product. Formally, we therefore
have the same vacuum \be \Phi_0={\langle z|0\rangle}=1\ee but we
stress again that the variables (and also the inner
product) are different, so this is not an equivalent vacuum. Particle states are
still monomials in the new variables: \be\Phi_n={\langle
z|n\rangle}\propto (G_{r{\cal P}}^\dagger)^n \Psi_0= z_{r{\cal P}}^n\ee but now
the non-normalizable states are the non-physical modes ${\cal P}={\cal P_-}=-1$.

The physical Hamiltonian is therefore:
 \be {\cal H}^{ph}_{eff}\approx
\frac{1}{2l_P^2}\int d\vk \sum_r   [{G}^{ph}_{r}{ G}^{ph
\dagger}_{r} (1+ir\gamma)+ {G}_{r}^{ph \dagger}{G}^{ph}_{r}
(1-ir\gamma)]\nonumber \ee where ${G}^{ph}_{r}=G_{r {\cal P_+}}$.
We can see that after normal ordering, right and
left particles are exactly symmetric, but a chiral vacuum energy
$V_r$ is found with: \be\label{chiraleq}
\frac{V_R-V_L}{V_R+V_L}=i\gamma\; . \ee Strictly speaking this
calculation only covers imaginary $\gamma$. If $|\gamma|>1$, the
vacuum energy of one of the modes becomes negative. This may
signal underlying fermionic degrees of freedom~\cite{cooper}. It
also implies that the calculation cannot be valid for the power
spectrum, a matter on which we now comment.

\section{Vacuum energy vs. fluctuations and ordering issues}\label{vacfluc}
We conclude with two final refinements to our calculation.
Firstly, we make the important distinction between vacuum energy
and vacuum fluctuations. It turns out that the chirality in these
two quantities is identical when $\gamma=\pm i$, but not
otherwise. Secondly we note that our results depend on the
ordering prescription. We comment on this dependence and explain
why it is a valuable asset.

\subsection{Vacuum energy and vacuum fluctuations}
\label{vac-en-flct} In the standard inflationary calculation, the
vacuum fluctuations (or their 2-point function) closely mimic the
vacuum energy. In both cases, it is important to compute the
time-dependent functions (\ref{psie}) multiplying creation and
annihilation operators. These provide the same spectrum for vacuum
energy and fluctuations, converting a $1/k$ spectrum inside the
horizon into a scale-invariant $1/k^3$ spectrum for $|k\eta|\ll 1$
for a deSitter background. Strictly speaking, however, the
chirality computed in Section~\ref{quantum ham} refers to the
vacuum energy. It turns out that the chiral asymmetry in the
vacuum energy and in its fluctuations is identical for the extreme
cases $\gamma=\pm i$, but not otherwise. This might have been
expected from the fact that for $|\gamma|>1$ one of the helicity
modes has negative vacuum energy. Its two-point function, being a
variance, must however always be positive.

We want to compute the power spectrum: \be \label{PS1}{\langle
0|A^\dagger_r(\vk)A_r(\vk')|0\rangle} =P_r(k)\delta(\vk-\vk')\;
,\ee where $A_r(\vk)$ is the Fourier component with handedness $r$
of the connection. Up to normalization factors this is given by:
\be\label{Bigak} A_r(\vk)=a_{r+}(\vk) e^{-i k\cdot x} +
a_{r-}^\dagger(\vk) e^{i k\cdot x} \ee before reality conditions
are imposed and the physical states selected. At this stage there
isn't any chirality in the 2-point function; it only creeps in
once the physical states are selected. For example, for $\gamma=i$
we find that \bea
A_R^{ph}(\vk)&=&a_{R+}(\vk) e^{-i k\cdot x} =g_{R+}(\vk) e^{-i k\cdot x} \\
A_L^{ph}(\vk)&=&a^\dagger_{L+}(\vk) e^{i k\cdot x} =
g^\dagger_{L+}(\vk) e^{i k\cdot x} \; ,\eea which creates a chiral
asymmetry (the right handed physical mode is represented by an
annihilation operator; the left-handed by a creation operator).
And indeed: \bea {\langle
0|A^{ph\dagger}_R(\vk)A^{ph}_R(\vk')|0\rangle}&=&
{\langle 0|g^\dagger_{R+}(\vk)g_{R+}(\vk)|0\rangle}=0\nonumber\\
{\langle 0|A^{ph\dagger}_L(\vk)A^{ph}_L(\vk')|0\rangle}&=&
{\langle 0|g_{L-}(\vk)g_{L-}^\dagger(\vk)|0\rangle}\neq0\nonumber
\eea
leading to vacuum fluctuations for left gravitons only.

For a general $\gamma$, Eqns.~(\ref{Gr+})-(\ref{Gr+dagger})
can be used to express the positive and negative frequencies of
the connection in terms of physical modes according to: \bea
a^{ph}_{r+}&=&\frac{r-i\gamma}{2r}G_{r{\cal P_+}}\\
a^{ph\dagger }_{r+}&=&\frac{r-i\gamma}{2r}G^\dagger_{r{\cal P_+}}\\
a^{ph}_{r-}&=&\frac{r+i\gamma}{2r}G_{r{\cal P_+}}\\
a^{ph\dagger}_{r-}&=&\frac{r+i\gamma}{2r}G^\dagger_{r{\cal P_+}}\;
. \eea Thus when we write (\ref{Bigak}) in terms of physical modes
we obtain: \bea A^{ph}_r(\vk)&=&\frac{r-i\gamma}{2r}G_{r{\cal
P_+}}(\vk) e^{-i k\cdot x} +
\frac{r+i\gamma}{2r} G_{r{\cal P_+}}^\dagger(\vk) e^{i k\cdot x} \nonumber \\
A_r^{ph\dagger}(\vk)&=&\frac{r+i\gamma}{2r}G_{r{\cal P_+}}(\vk)
e^{-i k\cdot x} + \frac{r-i\gamma}{2r} G_{r{\cal
P_+}}^\dagger(\vk) e^{i k\cdot x}\nonumber \eea so that \be
{\langle
0|A^{ph\dagger}_r(\vk)A^{ph}_r(\vk')|0\rangle}=\frac{(r+i\gamma)^2}{4}
{\langle 0|G_{r{\cal P_+}}(\vk)G^\dagger_{r{\cal
P_+}}(\vk)|0\rangle}\; . \ee This quantity is always positive for
imaginary $\gamma$, as it should be. We can now evaluate the
chiral asymmetry in the power spectrum, with the result:
\be\label{chiralP}
\frac{P_R-P_L}{P_R+P_L}=\frac{2i\gamma}{1-\gamma^2} \; .\ee This
is the counterpart to Eq. (\ref{chiraleq}), the asymmetry in the
vacuum energy. We see that the expressions only agree for
$\gamma=\pm i$, which maximize the chirality. The chirality
vanishes in the limits $|\gamma|\rightarrow 0$ and
$|\gamma|\rightarrow \infty$, the latter representing the
Palatini-Kibble theory.

\subsection{Ordering and experiment in quantum mechanics}
\label{order} Our results depend on the ordering prescription.
This is not a weakness peculiar to our paper or indeed to quantum
gravity: it is a general shortcoming of quantum mechanics. Many
quantum ordering issues are ultimately decided by experiment.
Failing that, there's prejudice. Arguments have been put forward
in favor of ``$E$s to the right of $A$s'' (it leads to simpler
expressions in the representation diagonalizing the connection),
but support for the opposite ordering is now ubiquitous in the
literature. For the Hamiltonian constraint these orderings
translate into $FEE$ or $EEF$ ordering, but ``$EFE$'' and
symmetric orderings are also popular, with the $EEKK$ term
(present if $\gamma\neq\pm i$) always symmetrized.

Theoretical arguments may be interesting and even useful, but as
with {\it any} other quantum mechanical description, ultimately
one must appeal to experiment to settle the
matter~\footnote{Notice that the requirement of hermiticity, much
used in more ``down to earth'' atomic and condensed matter
physics, is of no use here, since the inner product is not known a
priori. Even when hermiticity can be appealed to ambiguities creep
in, e.g. in atomic physics.}. It is in this spirit that we
consider the dependence of our results upon ordering as an asset
rather than a liability. Our results can assist us in resolving
ordering issues via experiment.

Consider a quantum version of the Hamiltonian (\ref{fullH}), with
symmetric $EEKK$ term, and the following general ordering for the
other term: \be EEF\rightarrow \alpha EEF+\beta FEE +\delta EFE\ee
with $\alpha+\beta+\delta=1$. It is easy to adapt the calculation
in Section~\ref{quantum ham} to find that the counterpart of
(\ref{chiraleq}) is now: \be\label{chiraleq2}
\frac{V_R-V_L}{V_R+V_L}=i\gamma (\alpha-\beta)\; . \ee

The asymmetry in the 2-point function is subject to a similar ordering
issue.
We can replace (\ref{PS1}) by: \be A^\dagger A \rightarrow
\epsilon A^\dagger A + \zeta A A^\dagger\; , \ee with
$\epsilon+\zeta=1$ and $\epsilon,\zeta>0$. This leads to \be
\label{chiralPR}\frac{P_R-P_L}{P_R+P_L}=\frac{2(\epsilon
-\zeta)i\gamma}{1-\gamma^2}\; , \ee generalizing (\ref{chiralP}).
In principle the parameters $\{\alpha,\beta,\delta\}$ can be
independent of $\{\epsilon,\zeta\}$, so we have derived
independent expressions. In a future publication~\cite{vacuum},
however, we
shall explain how the ordering of the Hamiltonian and the
two-point function are related, once one proposes a concrete wave
function for the ground state of the theory.

What are the phenomenological implications of this result? For a fuller
discussion we refer the reader to~\cite{TBpapers}. We want to stress, 
however, that 
the toy model employed there is in no way related to the 
calculation presented in this paper, but a translation is easy to carry out.
We note that expression (\ref{chiralPR}) is the only relevant input into the 
calculation of the TB and EB components of the polarization, as detailed
in~\cite{TBpapers}. Specifically, adapting the calculation in~\cite{TBpapers}
leads to:
\be
\frac{C_2^{TB}}{C_2^{BB}}\approx 800 \frac{(\epsilon
-\zeta)i\gamma}{1-\gamma^2}\; , 
\ee
for the ratio of tensor induced TB and BB quadrupole modes. This is 
an interesting quantity to consider, since it allows us to quantify 
how much easier the detection of a gravitational 
wave background (should it exist) would be rendered by chirality, regardless 
of the precise details of the model. 
The implication is that for a standard 
``extreme'' ordering (e.g. A's to the left of E's) a TB measurement would be
larger than a BB signal for an imaginary Immirzi parameter in the rough range
\be\label{bound}
\frac{1}{800}<|\gamma|<800\; .
\ee 
However, any direct constraints on the Immirzi parameter from current bounds
will necessarily be intertwined with parameter $r$ (the tensor to scalar 
ratio) and therefore be very model dependent. For example, if $r=0$, 
obviously no constraint can arise. Our result is therefore more useful
under the prospect of a positive detection of tensor modes in general
(cf.~\cite{TBpapers}).

We should not be surprised by the power of the prediction in Eq.~(\ref{bound}):
TB correlates something large with something small
rather than two small quantities, as is the case with BB. Therefore
even modest amounts of chirality
would render TB the method of choice for detecting tensor modes.

\section{Conclusions}\label{concs}
In this paper we provided the details behind an earlier
Letter~\cite{prl}, where it was claimed that a re-examination of
the inflationary mechanism for producing tensor modes within
Ashtekar gravity would render them chiral. Our efforts were
twofold: classical and quantum. Classically we ``rediscovered''
standard cosmological perturbation theory within Ashtekar's
formalism (Sections~\ref{classical}, \ref{realitytorsion}
and~\ref{secham}). The exercise proved far from trivial and
provided the following insights.
\begin{itemize}
\item {\bf The problem of time and the Hamiltonian constraint}
Time evolution in General Relativity is a diffeormorphism. Since
the theory is diffeormorphism invariant, time evolution reduces to
a constraint: the Hamiltonian constraint. This leads to the
problem of time---or of the lack thereof---in quantum gravity.
Perturbation theory provides an insight on how time might appear
as an illusion within a theory which has none. Perturbations arise
in a Russian doll scenario. Expanding the variables into first
order, second order, and so on, we find that structures, such as
the Hamiltonian constraint, exhibit an inter-locked structure
within the perturbative expansion (see Section~\ref{sechamconst}).
For example, the second order Hamiltonian constraint ${}^2{\cal H}$
is made up of terms quadratic in first order variables,
${}^2_1{\cal H}$, and terms linear in second order variables,
${}^2_2{\cal H}$. If all we care about are the first order
variables (such as in cosmological perturbation theory) then only
the first type of terms matter and the Hamiltonian constraint does
not apply to them. The diffeormorphism invariance of full GR is
then broken and we perceive a dynamical time. It is the terms
linear in the second order variables, usually called the
backreaction, that enforce diffeomorphism invariance to second
order, but these are ignored perturbatively. Time is an illusion
of perturbation theory, and yet that's the set up we live in.

\item {\bf Perturbation variables as a canonical transformation}
If one expands the full Ashtekar variables using perturbation variables
used by cosmologists and naively evaluates ${}^2_1{\cal H}$, then,
contrary to expectations, one does {\it not} obtain the standard
cosmological Hamiltonian (see Section~\ref{secham}). The mystery
is solved by regarding the perturbative expansions (\ref{eexp})
and (\ref{aexp}) as a canonical transformation into new variables,
$a^i_a$ and $\delta e^a_i$, which happen to be ``small'' (see
Section~\ref{twosubtle}). Thus the matter is far from pedantic,
but it also gives a more rigorous meaning to the perturbative
quantization procedure. We are not quantizing the fluctuations
whilst ``freezing'' the quantum mechanics of the background; we
are merely quantizing the full theory in new variables, which
happen to be ``small'' in some circumstances.

\item {\bf Boundary term} The boundary term described by equation
(\ref{boundary}) has been often ignored in the literature, usually
invoking suitable fall-off conditions~\cite{gravitons,btashtekar}.
In Section~\ref{twosubtle} we saw that this leads to the wrong
result for the cosmological Hamiltonian. The reason is that planes
waves, the central tool of cosmological perturbation theory, do
not satisfy the fall-off conditions, say, in a deSitter
background. Therefore the boundary term has to be included in
order to obtain the correct Hamiltonian to be employed in
quantizing the graviton modes (Section~\ref{quantum ham}).

\item {\bf Torsion and the Gauss constraint} Solving for the
perturbations to first order allows for a non-vanishing torsion
and Gauss constraint to second order (in the form of expressions
quadratic in linear perturbation variables). To use the notations
defined in the text, ${}^2_1G_i\neq 0$ and ${}^2_1T^a\neq 0$. As
with the Hamiltonian constraint, it is the second order (or
backreaction) terms ${}^2_2G_i$ and ${}^2_2T^a$, that enforce
these constraints to second order. In deriving the Ashtekar
Hamiltonian from the ADM formalism~\cite{thbook} these two
constraints are used, and therefore ${}^2_1{\cal H}$ acquires
extra terms. However, these turn out to be irrelevant full
divergences (see Section~\ref{sechamconst}).
\end{itemize}

In the second part of the paper (Sections~\ref{coms},~\ref{quantum
ham} and~\ref{vacfluc}) we then quantized this Hamiltonian theory.
Our work corrects a number of deficiencies present in previous
efforts~\cite{gravitons,leelaur}, and our improvements can be
traced to expansions (\ref{fourrier}). These always assign the
correct direction of motion and polarization to each mode, thereby
removing spurious couplings between $\vk$ and $-\vk$ modes, as
explained in Section~\ref{classical}. The commutation relations
were identified in Section~\ref{coms}, where it was noted that the
theory is very different from a complex scalar field, since its
Hamiltonian is complex off-shell. The quantum Hamiltonian was
evaluated in Section~\ref{quantum ham}, and it was used to read
off creation and annihilation operators for the graviton; these
are chirally described in terms of metric and connection
operators. In this paper we explained in detail the calculation
outlined in~\cite{prl}, and extended it for $\gamma\neq \pm i$. In
all cases we mimic Witten's result~\cite{wittenym}, initially
devised for Yang-Mills theory: half the graviton modes have
negative energies. However this is before reality conditions are
imposed, as made evident by the fact that we have twice as many
modes as needed: gravitons and anti-gravitons with right and left
polarizations. Upon determination of the inner product
(representing the reality conditions) we find that half of these
modes are not physical and these are the modes which have negative
energies in the quantum theory (and which also can be found not to
satisfy the classical equations of motions). Therefore only
non-pathological modes survive once the inner product and the
reality conditions are taken into account. A major novelty
emerges, however. We find that the physical modes have a chiral
representation in terms of metric and connection operators, the
exact expression depending on Immirzi parameter $\gamma$. As a
result the spectrum of gravitons is unchanged, but the vacuum
fluctuations and energy are chiral, with obvious cosmological
consequences~\cite{TBpapers}. In Section~\ref{vacfluc} we spelled
out how the exact expression depends not only on $\gamma$ but also
on the ordering prescription being used.

What is the physical origin of our main result? As we stressed in
Section~\ref{sechamconst} the Hamiltonian of quantum gravity is
intrinsically complex. It only becomes real on-shell, because the
Hamiltonian constraint forces it to vanish. However, when we
identify the terms that drive the perturbative gravitons,
${^{2}_1{\cal H}}$, we find that this constraint is waived, as
explained in Section~\ref{sechamconst}. Then, we have to face
the fact that for an imaginary (or more generally complex)
$\gamma$, the Hamiltonian is complex, which is ultimately the root
of all the new results reported in this paper. Many oddities found
in Sections~\ref{coms},~\ref{quantum ham} and~\ref{vacfluc} can be
directly traced to this fact, as highlighted in the text. We do
stress, however, that we never lose Hermiticity. In fact the
Hamiltonian is Hermitian with respect to the inner product used to
implement the reality conditions (Section~\ref{quantum ham}). In
this respect quantum gravity is very similar to the
``non-Hermitian'' Hamiltonians studied by Bender and
collaborators~\cite{benderetal}. These might be regarded as
nothing more than complex Hamiltonians which are in fact Hermitian
with respect to a non-trivial inner product.

Whilst the considerations in the previous paragraph are an important
contribution to understanding the physical origin of our results,
there probably exists an underlying deeper reason. One possibility presents itself by noting 
that the Immirzi parameter is associated with a surface or topological term 
(see, for example, Eqn.~(\ref{holst})). 
This suggests that the new effects may be due to 
instanton 
fluctuations\footnote{We thank the referee for bringing this possibility
to our attention.}. However this is far from obvious in our calculation,
where $\gamma$ is more readily understood as a parameter related to a canonical 
transformation. Consequently, bridging our results with those of~\cite{merc}
is far from straightforward (for example, how do the graviton operators 
derived in Section~\ref{quantum ham} fit in with the transition amplitudes 
of~\cite{merc}?). Nonetheless this remains a very interesting conjecture
and should be the subject of further investigation.

In future work~\cite{vacuum} we hope to shed more light on our
results by evaluating the wave function representing the vacuum
state identified in this paper. Obviously we have derived a range
of different vacua, one for each value of $\gamma$. A major result,
to be presented in~\cite{vacuum}, but which we wish to highlight
here, is that the vacua of the second quantized perturbation
theories examined in this paper are {\it never} the perturbed
Kodama state, which therefore can never describe standard
gravitons~\cite{leelaur}. The reason rests on a very simple and
clear-cut algebraic fact. The implication is very deep. It appears
that taking the semi-classical limit of quantum gravity is never
the correct path for a perturbative, but fully quantized 
theory. The perturbed semi-classical limit is in fact at odds with
the full quantization of the perturbations. We will expand on this
fact making contact with previously published 
work on other vacuum states~\cite{cooper}.
We'll also explain the relation with previous work on the graviton
propagator~\cite{recentgravitons}, where a chiral contribution was found. 
The relation with our results is not obvious,
since~\cite{recentgravitons} employed an Euclidean signature and a
real $\gamma$. However a modification of their work can be
explained within our framework.

In this paper we have restricted ourselves to an imaginary Immirzi 
parameter. Throughout we have flagged a number of places
where this assumption was used. Relaxing it renders 
several formulae rather cumbersome, and the algebra more involved.
These changes will be presented in a future publication~\cite{future}, where 
we'll also explain a number of further subtleties
that arise if $\gamma$ is allowed to be real or have a real part. Nonetheless, the
final result is, remarkably, very simple. After a lengthy
calculation it turns out that for a complex $\gamma$ the chirality formulae 
presented in this paper are trivially changed by replacing $\gamma$ by 
its imaginary part. An immediate implication is that  according to our 
calculations the real theory has no chirality at all. A full discussion 
of the implications will be presented in~\cite{future}.

On a more cosmological front, another issue we'd like to revisit
is the behavior of the modes outside the horizon. Inspection of
the Hamiltonian reveals that the dynamics then becomes fully
driven by the metric, with the connection pushed aside. Thus, whereas the
Ashtekar formalism may be expected to result in novelties inside
the horizon, a complete reduction to the second order formalism is
expected as the modes leave the horizon. The problem of
decoherence in cosmology may benefit from this insight. It seems
that at the same time as the quantum modes lose their phase and
prepare themselves to ``go classical'', the
quantum distinction between metric and connection also evaporates.

In the meantime we have shown how a perturbative re-examination of
quantum gravity can be fruitful. We hope to have cleared up a few
misconceptions and paradoxes. Above all, we derived a striking
prediction for the theory, which could be tested in upcoming CMB
polarization experiments. There are other mechanisms to generate
gravitational chirality (e.g.~\cite{steph,cooper,merc}), but the
one pointed out in this paper is by far the simplest. As stressed
in~\cite{TBpapers}, even modest chirality in the gravitational
wave background would render its detection far easier, and probably
within the reach of the PLANCK mission.

\begin{acknowledgments}
We thank Martin Bojowald, Gianluca Calcagni, Carl Bender,
Jonathan Halliwell, Chris Isham, Kirill Krasnov, Hermann Nicolai,
Lee Smolin, Julian Sonner, Kelly Stelle and Thomas Thiemann for
help regarding this project.
\end{acknowledgments}

\section*{Appendix I: A precis of cosmological perturbation theory}
Here we briefly collect the relevant results of cosmological tensor
perturbation theory (see for example~\cite{lid,muk}). As in the main
text we start from:
\be
ds^2=a^2(-d\eta^2 +(\delta_{ij}+h_{ij})dx^idx^j)
\ee
where $h_{ij}$ is divergenceless and tracefree.
Then (raising an lowering the indices of $h_{ij}$ with $\delta_{ij}$),
the second order Einstein-Hilbert action is:
\begin{equation}\label{haction}
S=\frac {1}{64 \pi G}\int a^2 (h'^{ij} h'_{ij} - h^{ij,k}
h_{ij,k}) \; d^3x d\eta,
\end{equation}
where $'$ denotes derivative with respect to $\eta$. Expanding in
Fourier modes, and writing $h_{ij}(\vk,\eta)=
h(\vk,\eta)\epsilon_{ij}(\vk)$ (where $\epsilon_{ij}$  is the
polarization tensor) we can then define a ``$v$'' variable
(similar to that used for treating scalar fluctuations) by: \be
v(\vk,\eta)={\sqrt{\frac{\epsilon^{ij}{\overline\epsilon}_{ij}}{32\pi
G}}} a h(\vk,\eta) \ee so that the action becomes: \be
S=\frac{1}{2}\int d^3k\, d\eta\,
{\left(v'^2-\left(k^2-\frac{a''}{a}\right)v \right)}\; . \ee This
is identical to the action used in the treatment of scalar
fluctuations. Its associated Hamiltonian is: \be\label{cosmoH}
H=\frac{1}{2}\int d^3k\,
{\left(v'^2+\left(k^2-\frac{a''}{a}\right)v \right)}\; . \ee In a
deSitter universe it leads to the equation of motion: \be\label{veq}
v''+{\left(k^2-\frac{2}{\eta^2}\right)}v=0\; . \ee This can be
generally solved with Bessel functions, but for deSitter the
solution is very simple.

We should impose the boundary condition for $k|\eta|\gg 1$:
\be
v\rightarrow \frac{e^{-ik \eta}}{\sqrt {2 k} } b
\ee
so that upon second quantization amplitudes $b$ and $b^\dagger$ become
annihilation and creation operators. The vacuum expectation value
can then be evaluated and followed for modes outside the horizon
($k|\eta|\ll 1$), obtaining scale-invariance. The full solution
for deSitter is:
\be
v=\frac{e^{-ik \eta}}{\sqrt {2 k} } {\left( 1-\frac{i}{k\eta}
\right)}b
\ee
so that in the limit $k\eta\ll 1$ we have
\be
{\langle 0||v(k)|^2|0\rangle}=\frac{1}{2^{\frac{3}{2}}k^3\eta}\; .
\ee
The reason variable $v$ is chosen instead of $h_{ij}$ is to get rid
of the friction term in
\be
h_{ij}''+2\frac{a'}{a}h_{ij}'+k^2h_{ij}=0\; .
\ee

\section*{Appendix II: Derivation of the generating function}

Following \cite{goldstein}, the transformation between two sets of
canonical variables is achieved by a generating function that
links the different variables. If the canonical transformation is
time dependent, the Hamiltonian will change. In our case, we have
a generating function $F(A^i_a, \delta e^a_i,\eta)$ that can be
can be determined from \be \label{F1} \frac{\delta F}{\delta
A^i_a}=\frac{1}{\gamma l_P^2}E^a_i \ee \be \label{F2} \frac{\delta
F}{\delta \left(\delta e^a_i \right)}=-\frac{1}{\gamma
l_P^2}a^i_a\; . \ee The new Hamiltonian ${\cal K}$ is related to
the old one by \be \label{Hchange} {\cal K}\left(a^i_a,\delta
e^a_i \right)={\cal H}\left(a^i_a,\delta e^a_i
\right)+\frac{\partial F}{\partial \eta}\left(a^i_a,\delta e^a_i
\right) \ee where ${\cal H}$ and $\frac{\partial F}{\partial
\eta}$ have to be expressed in terms of the new variables. Solving
equations (\ref{F1}) and (\ref{F2}) yields \be F(A^i_a, \delta
e^a_i,\eta)=\int d^3 x \frac{1}{\gamma l_P^2} \left(a^2 A^i_i -a
\delta e^a_i A^i_a\right) \ee Thus, \be \frac{\partial F}{\partial
\eta}=\int d^3 x \frac{1}{\gamma l_P^2} \left(2aa' A^i_i -a'
\delta e^a_i A^i_a \right) \ee Expressing this in terms of the new
variables (and ignoring a zeroth order contribution) we find \be
\frac{\partial F}{\partial \eta}(a^i_a, \delta e^a_i,\eta)=\int
d^3 x \frac{-1}{\gamma l_P^2} Ha \delta e^a_i a^i_a \ee Using
equation (\ref{Hchange}), the corrected Hamiltonian is therefore
\bea\label{effectham2} {\cal H} _{eff}&=&\frac{1}{2l_P^2}\int
d^3x\Bigl[ -a_{ij}a_{ij} -2 \epsilon_{ijk}
(\partial_j\delta e_{li}) a_{kl} \Bigr.\nonumber\\
&& \Bigl. -2 \left(\gamma+\frac{1}{\gamma} \right) Ha \delta e_{ij}a_{ij}
-\left(\gamma^2+3\right)H^2 a^2 \delta e_{ij} \delta e_{ij}
\Bigr]\;.\nonumber \\
\eea
\begin{widetext}
\section*{Appendix III: General form of the off-shell Hamiltonian
in terms of metric and connection variables}

For general $\gamma$, the Hamiltonian in terms of Fourier modes can be found by substituting the expansions (\ref{fourrier}) into the perturbative Hamiltonian (\ref{totalham}):

\bea\label{hamexp3} {\cal H}_{eff}&=&\frac{1}{l_P^2}\int d^3k\sum_r \frac{1}{\gamma^2} \biggl\{ \biggr.
\nonumber\\
&&
\Bigl[\bigl\{k^2\left(\gamma^2+1\right)-2\gamma^2 H^2 a^2 \bigr\} {\tilde e}_{r+}(\vk)-kr\left(\gamma^2+1\right) {\tilde a}_{r+}(\vk) \Bigr] {\tilde e}_{r+}(-\vk)
\nonumber\\
&+&
\Bigl[\bigl\{k^2\left(\gamma^2+1\right)-2\gamma^2 H^2 a^2 \bigr\} {\tilde e}_{r+}(\vk)-kr\left(\gamma^2+1\right) {\tilde a}_{r+}(\vk)\Bigr] {\tilde e}_{r-}^\dagger(\vk)
\nonumber\\
&+&
\Bigl[\bigl\{k^2\left(\gamma^2+1\right)-2\gamma^2 H^2 a^2 \bigr\}{\tilde e}_{r-}^\dagger(\vk)-kr\left(\gamma^2+1\right) {\tilde a}_{r-}^\dagger(\vk) \Bigr]{\tilde e}_{r+}(\vk)
\nonumber\\
&+&
\Bigl[\bigl\{k^2\left(\gamma^2+1\right)-2\gamma^2 H^2 a^2 \bigr\} {\tilde e}_{r-}^\dagger(\vk)-kr\left(\gamma^2+1\right) {\tilde a}_{r-}^\dagger(\vk) \Bigr]{\tilde e}_{r-}^\dagger(-\vk)
\nonumber\\
&+&
\Bigl[kr\left(\gamma^2-1\right){\tilde e}_{r+}(\vk) + {\tilde a}_{r+}(\vk) \Bigr]{\tilde a}_{r+}(-\vk)
\nonumber\\
&+&
\Bigl[kr\left(\gamma^2-1\right){\tilde e}_{r+}(\vk) + {\tilde a}_{r+}(\vk) \Bigr]{\tilde a}_{r-}^\dagger(\vk)
\nonumber\\
&+&
\Bigl[kr\left(\gamma^2-1\right){\tilde e}_{r-}^\dagger(\vk) + {\tilde a}_{r-}^\dagger(\vk) \Bigr]{\tilde a}_{r+}(\vk)
\nonumber\\
&+&
\biggl.\Bigl[kr\left(\gamma^2-1\right){\tilde e}_{r-}^\dagger(\vk) + {\tilde a}_{r-}^\dagger(\vk) \Bigr]{\tilde a}_{r-}^\dagger(-\vk) \biggr\} \;.\nonumber\\ \eea
\end{widetext}


\end{document}